 \documentclass[aps,prstab,a4paper,reprint,superscriptaddress, nobibnotes, longbibliography]{revtex4-1}
\usepackage{helvet}
\usepackage{graphicx}
\usepackage{xspace}
\usepackage{amsmath,amssymb,bm}
\usepackage{braket}
\usepackage{resizegather}
\usepackage{enumitem}
\usepackage[usenames,dvipsnames]{xcolor}
\usepackage{booktabs}
\usepackage{epstopdf}
\usepackage[above]{placeins}

\usepackage{MnSymbol}
\usepackage[colorlinks,urlcolor=blue,citecolor=blue,linkcolor=blue,
			pdfauthor={Ikonen, Salmilehto, Möttönen},
            pdftitle={Energy-Efficient Quantum Computing},
            pdfsubject={Greatest lower bound of single-qubit gate error. Protocol to reduce energy consumption of a quantum processor.},
            pdfkeywords={Qubit; Quantum gate; Quantum processor}]{hyperref}

\newif\ifdrafttext
\drafttextfalse
\ifdrafttext \usepackage[colorlinks,urlcolor=black,citecolor=black,linkcolor=black]{hyperref} \else    \fi

\makeatletter
\@addtoreset{figure}{myfigure}
\def\@caption@fignum@sep{.~}%
\def\fnum@figure{\figurename~\thefigure}
\makeatother

\makeatletter
\@addtoreset{table}{mytable}
\def\@caption@tablenum@sep{.~}%
\def\tnum@table{\textbf{\tablename~\thetable}}
\makeatother

\global\long\def\kb#1#2{|#1\rangle\!\langle#2|}
\global\long\def\e#1{\mathrm{e}^{#1}}
\global\long\def\m#1{\mathrm{#1}}
\global\long\def\err{\mathcal{E}}
\global\long\def\n{\overline{n}}
\global\long\def\abs#1{\left|#1\right|}
\global\long\def\sq{\ln\sqrt{\pi/2}}
\global\long\def\c{\hat{\chi}}
\global\long\def\s{\hat{\sigma}}
\global\long\def\k#1{\Ket{#1}}
\global\long\def\b#1{\Bra{#1}}

\ifdrafttext
\else

\fi

\newcommand{\titleofpaper}{Energy-Efficient Quantum Computing}
\newcommand{\QCDaffiliation}{QCD Labs, COMP Centre of Excellence, Department of Applied Physics, Aalto University, P.O. Box 13500, FI--00076 Aalto, Finland}

\newcommand{\Yaleaffiliation}{Department of Physics, Yale University, New Haven, Connecticut 06520, USA.}

\newcommand{\Jyvaaffiliation}{University of Jyv\"askyl\"a,
Department of Mathematical Information Technology, P.O. Box 35, FI-40014 University of Jyv\"askyl\"a, Finland.}

\begin{document}
%\linenumbers

\title{\titleofpaper}

\author{Joni Ikonen}
\email{joni.2.ikonen@aalto.fi}
\affiliation{\QCDaffiliation}

\author{Juha Salmilehto}
\affiliation{\Yaleaffiliation}

\author{Mikko M\"ott\"onen}
%\thanks{mikko.mottonen@aalto.fi}
\affiliation{\QCDaffiliation}
\affiliation{\Jyvaaffiliation}

\date{\today}
\vspace*{-1.1 cm}
\begin{abstract}
\noindent In the near future, a major challenge in quantum computing is to scale up robust qubit prototypes to practical problem sizes and to implement comprehensive error correction for computational precision.
Due to inevitable quantum uncertainties in resonant control pulses, increasing the precision of quantum gates comes with the expense of increased energy consumption.
Consequently, the power dissipated in the vicinity of the processor in a well-working large-scale quantum computer seems unacceptably large in typical systems requiring low operation temperatures.
Here, we introduce a method for qubit driving and show that it serves to decrease the single-qubit gate error without increasing the average power dissipated per gate.
Previously, single-qubit gate error induced by a bosonic drive mode has been considered to be inversely proportional to the energy of the control pulse, but we circumvent this bound by reusing and correcting itinerant control pulses.
Thus our work suggests that heat dissipation does not pose a fundamental limitation, but a necessary practical challenge in future implementations of large-scale quantum computers.

\end{abstract}
%x words


\maketitle

\newpage

\section{Introduction\label{intro}}

\noindent Quantum bits, or qubits~\cite{Preskill}, have been realized using, for example, superconducting circuits~\cite{Nakamura,Barends,Kelly}, quantum dots~\cite{Bonadeo,Veldhorst}, trapped ions~\cite{Sackett,Debnath}, single dopants in silicon~\cite{Pla}, and nitrogen vacancy centres~\cite{Togan}. The state of a qubit is affected by various sources of error such as finite qubit lifetime, measurement imperfections, non-ideal initialization, and imprecise external control. Provided that these errors are below a certain threshold, they can be corrected with quantum error correction codes~\cite{Terhal,Fowler,Kelly} which encode the information of a logical qubit into an ensemble of physical qubits. Surface codes~\cite{Fowler}, error correction codes with the highest known thresholds, may require thousands of physical qubits for each fault-tolerant logical qubit. Controlling such a large ensemble of qubits consumes a great amount of power, rendering heat management at the qubit register an important challenge.

The power consumption of a quantum processor can be decreased by implementing more accurate physical qubits, thus leading to smaller ensembles forming the logical qubits. However, it is known that gate errors also arise from the quantum-mechanical uncertainties in the control pulse~\cite{Ozawa0,Gea-Banacloche, Ozawa1,Ozawa2,Miller,Karasawa, Igeta}. In the case of a resonant disposable control pulse, this type of error is inversely proportional to the pulse energy, and hence poses a trade-off in the power management of the quantum computer. Even in the absence of all other types of error, this result implies such a high level of dissipated power at the chip temperature that it challenges the commercially available cryogenic equipment, as we estimate in Appendix~\ref{appA} for a typical superconducting quantum computer running a surface code to factorize a 2000-bit integer.

In this work, we derive the greatest lower bound for the gate error within the resonant Jaynes--Cummings model~\cite{Jaynes,Shore}. The inevitable error originates from the quantum nature of the driving mode and becomes dominant in the regime of low driving powers.  In contrast to previous work~\cite{Ozawa1, Ozawa2, Igeta}, our constructive derivation does not need to assume any particular state of the system and is applicable to qubit rotations of arbitrary  angles. In addition to the lower bound itself, our method naturally finds the bosonic quantum states of the pulse that reach the bound. We explicitly show that single-qubit rotations are optimally realized by applying a certain amount of squeezing to coherent states.

The optimal states do not alone solve the above-mentioned heat dissipation problem, but we additionally find that back-action-induced correlations between the control pulse and the controlled qubit can be transferred to auxiliary qubits (see also Refs. \cite{Layden,Slosser,Aberg}). Thus we propose a control protocol where multiple gates are generated with a single control pulse which is frequently refreshed using auxiliary qubits. Whereas previous studies suggest that it is not possible to save energy by reusing control pulses without sacrificing the minimum gate fidelity~\cite{Ozawa2}, our method exhibits orders of magnitude smaller energy consumption with no drop in the average gate fidelity.

This paper is organized as follows. In Sec.~\ref{sec:semiclassical}, we briefly summarize the formalism used to describe qubit rotations and discuss gate errors in the semiclassical model. In Sec.~\ref{sec:optimization}, we derive the quantum limit of gate error. The refreshing protocol is constructed and studied in Sec.~\ref{sec:protocol} and the key results are summarized and discussed further in Sec.~\ref{sec:discussion}.

\section{Semiclassical model\label{sec:semiclassical}}

\begin{figure}[tb] \center	
	\includegraphics[width=1\linewidth]{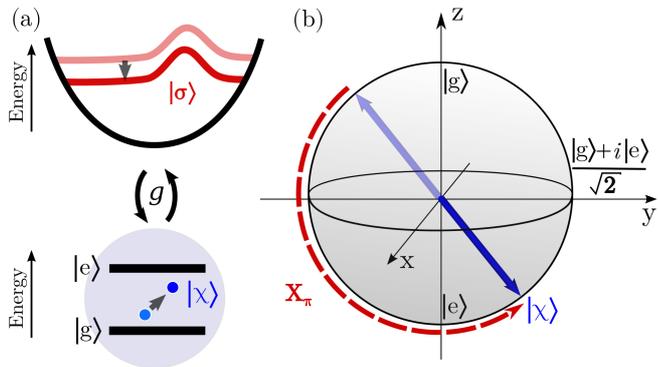}
	\caption{ \label{fig1} Model system. (a) Ideal two-level system (bottom) interacting with a harmonic oscillator (top). (b) Bloch vector representation of the qubit state $|\chi\rangle$ and an example $X_{\pi}$ rotation.
}
\end{figure}

Let us first review the semiclassical formalism of single-qubit control and the resulting gate errors. The state of a qubit can be represented as a Bloch vector constrained inside a unit sphere, see Fig.~\ref{fig1}. Single-qubit logic gates $R_{\theta}$, realized using, e.g., microwave pulses, rotate the Bloch vector by $\theta$ about the axis $R$. Assuming that the control pulse is a classical waveform in resonance with the qubit transition energy $\hbar\omega$, the system may be described in the rotating frame using a semiclassical interaction Hamiltonian of the form~\cite{Nakahara}
\begin{equation}
\hat{H}_{\m{\m{int}}}^{\m{cl}}(t)=\hbar g(t)\left(\alpha\kb{\m e}{\m g}+\alpha^{*}\kb{\m g}{\m e}\right), \label{eq:ClassicalH}
\end{equation}
where $\Ket{\m g}$ and $\Ket{\m e}$ denote the ground and excited states of the qubit, respectively, $\alpha=\abs{\alpha}\e{i\phi}$ represents the classical amplitude $\abs{\alpha}$ and phase $\phi$ of the control field, $g(t)$ is the coupling constant including the pulse envelope, and $\hbar$ is the reduced Planck constant. The gate $R_{\theta}$ is implemented by choosing the interaction time $T$ and the pulse envelope such that they satisfy $2\abs{\alpha}\int_{0}^{T}g(t)\m dt=\theta$. For example, setting $\theta=\pi$ and $R$ along the $x$-axis, the temporal evolution operator $\hat{U}_{\m{cl}}=\exp\left[-i\int_{0}^{T}\hat{H}_{\m{int}}^{\m{cl}}(t)\m dt/\hbar\right]$ becomes $\hat{U}_{\m{cl}}=-i\hat{\sigma}_{\m x}$, where $\hat{\sigma}_{\m x}=\kb{\m e}{\m g}+\kb{\m g}{\m e}$ is the Pauli $X$-operator. Thus, up to a redundant global phase factor, the interaction implements a perfect NOT gate $X_{\pi}$.

We assess gate errors by utilizing the state transformation error
\begin{equation}
\err_{\m{cl}}(\vartheta,\varphi)=1-\abs{\Bra{\chi_{0}}\hat{K}^{\dagger}\hat{U}{}_{\m{cl}}\Ket{\chi_{0}}}^{2}, \label{eq:classicalerror}
\end{equation}
where the initial qubit state is given by $\Ket{\chi_{0}}=\cos\left(\frac{\vartheta}{2}\right)\Ket{\m g}+\sin\left(\frac{\vartheta}{2}\right)\e{i\varphi}\Ket{\m e}$ and $\hat{K}$ is the desired gate. In general, the qubit state is unknown during the computation, and therefore we choose not to restrict our analysis to any specific state. Instead, we study the average of a given error measure $\err_i$ over a uniform state distribution on the Bloch sphere, generally given by
\begin{equation}
\overline{\err}_i=\frac{1}{4\pi}\int_{0}^{\pi}\int_{0}^{2\pi}\err_i(\vartheta,\varphi)\sin\vartheta\,\m d\vartheta\,\m d\varphi. \label{eq:AverageError}
\end{equation}

Semiclassically, a source of gate error arises from uncertainties in the phase and the photon number $n$, which are, for small phase fluctuations, fundamentally bounded by quantum mechanics through the minimal uncertainty relation~\cite{Pegg} $\Delta n\Delta\phi=1/2$. Thus we consider a control pulse with an average of $\n=\abs{\alpha}^{2}$ photons and minimal uncertainties $\Delta n=\sqrt{\n}\e{-r}$ and $\Delta\phi=\e{r}/(2\sqrt{\n})$, where $r$ is a free squeezing parameter. These uncertainties carry on to the temporal evolution operator $\hat{U}_{\m{cl}}$, and we find from Eq.~\eqref{eq:AverageError} that the average gate error becomes inversely proportional to the photon number. For the $X_{\pi}$ gate for example, we obtain the average gate error $\overline{\err}_{\m{cl}}=(4\e{2r}+\pi^{2}\e{-2r})/(24\n)$ in the limit $\n\rightarrow\infty$. Interestingly, the error is minimized with a non-zero squeezing parameter $r=\ln\sqrt{\pi/2}$, a result also obtained in the full quantum treatment in Sec.~\ref{minimizationmethod}. An alternative qubit-independent error quantity is the maximum gate error given by $\err_{\max}=\max_{\vartheta,\varphi}\err(\vartheta,\varphi)$, which obeys a similar $1/\n$-dependence~\cite{Igeta,Gea-Banacloche}.

\section{Quantum limit of gate error\label{sec:optimization}}

Let us proceed to the full quantum treatment, where the gate operation arises from the quantum-mechanical interaction between the qubit and a single bosonic mode referred to as the drive.
Utilization of such quantum drive~\cite{Salmilehto} allows us to account for the changes in its state arising from the interaction with the qubit. In practice, qubits are also driven by propagating photons described by a continuum of modes, but such arrangements do not save energy in comparison to a well-controlled single mode. Hence our description below is expected to yield a fundamental lower bound for the energy needed for controlling a single qubit at a given fidelity.

In contrast to the semiclassical model, the evolution of the qubit is not unitary. After the interaction, the qubit state is extracted by taking a partial trace over the drive degrees of freedom as
\begin{equation}
\hat{\chi}(T)=\m{Tr}_{D}\left[\hat{U}(T)\hat{\rho}_0\hat{U}^{\dagger}(T)\right],
\end{equation}
where $\hat{\rho}_0$ and $\hat{U}(T)$ denote the arbitrary initial density operator and the evolution operator of the qubit--drive system, respectively. The error, or infidelity, between the target and the resulting qubit state is here defined as
\begin{equation}
\err\left[\hat{\chi}_{0},\hat{\chi}(T)\right]=1-\m{Tr}\left[ \hat{\chi}(T)\hat{K}\hat{\chi}_{0}\hat{K}^{\dagger}\right], \label{eq:errorTraces}
\end{equation}
which can be regarded as a generalization of Eq.~\eqref{eq:classicalerror}.

\subsection{Gate error in the Jaynes--Cummings model\label{quantumerror}}

The dynamics of the qubit--drive system is generally described by the Jaynes--Cummings Model~\cite{Jaynes,Shore}, which includes the rotating-wave approximation.  Assuming resonant interaction, the system is governed by the interaction Hamiltonian
\begin{equation}
\hat{H}_{\m{int}}=\hbar g(t)\left(\kb{\m e}{\m g}\otimes\hat{a}+\kb{\m g}{\m e}\otimes\hat{a}^{\dagger}\right), \label{eq:JCMHamilton}
\end{equation}
where $\hat{a}$ is the bosonic annihilation operator of the drive mode. Without loss of generality, we assume an on-off envelope such that $g(t)=\m{const.}$ for $0<t<T$ and $g(t)=0$ otherwise. Most features of the semiclassical model are reobtained if $\abs{\alpha}\rightarrow\infty$ and the drive is in the coherent state $\Ket{\alpha}=\e{-\frac{1}{2}|\alpha|^{2}}\sum_{n=0}^{\infty}\frac{\alpha^{n}}{\sqrt{n!}}\Ket{n}$, where $\Ket{n}$ is the $n$th Fock state. For example, taking the expectation value of $\hat{H}_{\mathrm{int}}$ in the state $\Ket{\alpha}$ yields the semiclassical Hamiltonian in Eq.~(\ref{eq:ClassicalH}). Thus the coherent state approximately induces a gate $R_{\theta}$ if the timing condition $2gT\abs{\alpha}=\theta$ is satisfied.

If the initial state of the joint system is separable, $\hat{\rho}_0=\kb{\chi_{0}}{\chi_{0}}\otimes\kb{\sigma_{0}}{\sigma_{0}}$, where $\k{\chi_{0}}$ and $\k{\sigma_{0}}$ denote the initial qubit and drive states, respectively, the gate error of Eq.~\eqref{eq:errorTraces}	induced by the Jaynes--Cummings interaction can be written in the general form
\begin{equation}
\err_{i}(\sigma_{0})=1-\Bra{\sigma_{0}}\hat{F}_{i}\Ket{\sigma_{0}}. \label{eq:generalCompact}
\end{equation}
Here, $\err_{i}$ denotes either the transformation error $\err(\vartheta,\varphi)$ of a particular qubit state, the average gate error $\overline{\err}$ [Eq.~\eqref{eq:AverageError}], or the maximum gate error $\err_{\max}$. The information about the desired gate and chosen interaction time is contained in the corresponding operator $\hat{F}_{i}$ which is denoted either by $\hat{F}(\vartheta,\varphi)$, $\hat{F}_\textrm{avg}$, or $\hat{F}_{\min}$, respectively. An analytical expression for $\hat{F}(\vartheta,\varphi)$ and $\hat{F}_\textrm{avg}$ can be found for any gate, whereas an expression for $\hat{F}_{\min}$ exists for at least rotations $R'_{\pi}$, where the rotation axis
$R'$ is restricted to the $xy$-plane of the Bloch sphere. See Appendix \ref{appB} for derivations and detailed expressions.

\subsection{Minimization of gate error\label{minimizationmethod}}

\begin{figure}[tb!] \center	
	\includegraphics[width=0.95\linewidth]{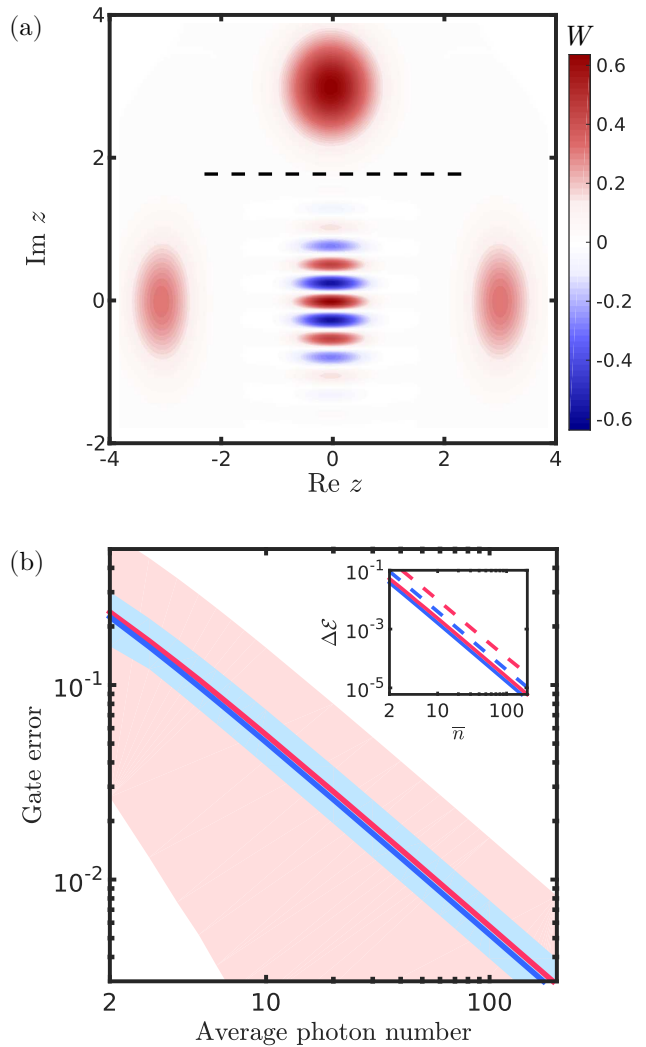}
	\caption{ \label{fig2} Optimal drive states and the resulting error.
	(a) Numerically solved initial drive states $\Ket{\sigma_{0}^{\m{opt,avg}}}$ that minimize the average error of rotations $Y{}_{\pi/2}$ and $X_{\pi}$ as Wigner distributions above and below the dashed line, respectively. The Wigner function is defined as $W(z)=\frac{2}{\pi}\m{Tr}_{D}\left[\hat{D}(-z)\kb{\sigma_{0}^{\m{opt}}}{\sigma_{0}^{\m{opt}}}\hat{D}(z)\e{i\pi\hat{a}^{\dagger}\hat{a}}\right]$,
where $\hat{D}(z)$ is the displacement operator. The interaction time for each operation $R_{\theta}$ is $T=\theta/(6g)$, which is expected to yield states with $\n=9$.
	(b) Gate error for an $X_{\pi}$ operation as a function of the average photon number $\n$ of the driving pulse which is initialized either in the coherent state  (red color) or the squeezed cat state (blue color). The highlighted areas indicate the range of error, depending on the initial state of the qubit, and the solid lines show the error averaged over qubit states distributed uniformly on the Bloch sphere, $\overline{\err}$. The inset shows the difference $\Delta\err$ between the numerically calculated errors and their analytical first-order approximations (Table~\ref{tab_1}), with dashed lines indicating the difference in maximum errors.
	}% end caption
\end{figure}

\begin{table*} [bt!]
		\centering
		\caption{\label{tab_1}Analytical expressions for gate errors using different drive states. Average and maximum errors for gates $R'_{\pi}$ and $R'_{\pi/2}$ to the first order in $\n\,^{-1}$, implemented with the drive pulse either in a coherent state, in a squeezed coherent state, or in a squeezed cat state. Here, $R'$ lies in the $xy$-plane forming an angle $\phi$ with respect to the $x$-axis. The squeezing parameter, $r$, is chosen in each case to minimize the average error $\overline{\err}$. The percentages denote the error in units of the lower bounds marked with ({*}).}
		
\renewcommand{\arraystretch}{2}%
\begin{tabular}{c|ccccc|ccc}
	\hline
 	{}	  & \multicolumn{5}{c|}{$R'_{\pi}$} & \multicolumn{3}{c}{$R'_{\pi/2}$}\tabularnewline
	\hline
	State  & \multicolumn{2}{c}{$\overline{\err}$} & \multicolumn{2}{c}{$\err_{\max}$} & $r$  & \multicolumn{2}{c}{$\overline{\err}$} &$r$  \\
	\hline
	\hline
	Coherent, $|\sqrt{\n}\e{i\phi}\rangle$ & $\frac{4+\pi^{2}}{24\n}$ & (110\%) & $\frac{4+4\pi+\pi^{2}}{16\n}$  & (210\%)  & 0  & $\frac{8+\pi^{2}}{96\n}$  & (100.5\%)& 0 \\
	\hline
	Squeezed, $|\sqrt{\n}\e{i\phi},r\rangle$  & $\frac{\pi}{6\n}$  & ({*})  & $\frac{\pi}{2\n}$ & (200\%) & $\e{2i\phi}\ln\sqrt{\pi/2}$  & $\frac{\sqrt{2}\pi}{24\n}$  & ({*})& $\e{2i\phi}\ln\sqrt{\pi/\sqrt{8}}$ \\
	\hline
	S. cat, $|\sqrt{\n}\e{i\phi},r\rangle\pm|-\sqrt{\n}\e{i\phi},r\rangle$  & $\frac{\pi}{6\n}$  & ({*})  & $\frac{\pi}{4\n}$  & ({*}) & $\e{2i\phi}\ln\sqrt{\pi/2}$  & \multicolumn{3}{c}{N/A} \\
	\hline
\end{tabular}
\end{table*}

We solve the drive states that minimize the average or maximum gate error for a given interaction time and a desired rotation $R'_\theta$. To this end, it is sufficient to consider only pure states~\cite{Nielsen}, and hence we may employ the forms given by Eq.~\eqref{eq:generalCompact}. The error-minimizing states are the eigenstates of operators $\hat{F}_{i}$ that correspond to the largest eigenvalue $f_{i}$,
\begin{equation}
\hat{F}_{i}\Ket{\sigma_{0}^{\m{opt},i}}=f_{i}\Ket{\sigma_{0}^{\m{opt},i}}.
\end{equation}
By definition, the optimal states $\Ket{\sigma_{0}^{\m{opt},i}}$ provide a fundamental lower bound for the error $\err_{i}$.% in the resonant Jaynes--Cummings model.

We solve this eigenvalue equation numerically. Examples of fidelity-optimal solutions are shown in Fig.~\ref{fig2}a using the Wigner pseudo-probability function~\cite{Dodonov}. The numerically obtained states can be accurately described using the squeezed coherent states $\Ket{\alpha,r}=\hat{D}(\alpha)\hat{S}(r)|0\rangle$, where $\hat{D}(\alpha)=\e{\alpha\hat{a}^{\dagger}-\alpha^{*}\hat{a}}$ and $\hat{S}(r)=\e{\frac{1}{2}r^{*}\hat{a}^{2}-\frac{1}{2}r\left(\hat{a}^{\dagger}\right)^{2}}$ are the displacement and squeezing operators, respectively~\cite{Dodonov}. Importantly, the numerical solutions possess the correct amplitude and phase to satisfy the timing condition $2gT\abs{\alpha}=\theta$ and to set the desired direction of the rotation axis, without imposing them explicitly. Furthermore, the average errors, as well as the optimal squeezing parameters, are equal to those obtained in the semiclassical approach in Sec.~\ref{sec:semiclassical}.

In the specific case of $\pi$-rotations, a sum of two eigenvectors, i.e., the squeezed cat state~\cite{Govia,Dodonov,Vlastakis}
\begin{equation}
\Ket{\Sigma(\alpha)}=\frac{1}{\mathcal{N}}\left(|\alpha,\sq\rangle\pm|-\alpha,\sq\rangle\right), \label{eq:Optimal state}
\end{equation}
where the positive constant $\mathcal{N}$ ensures normalization, is a state that minimizes both the average and the maximum error simultaneously (see Appendix~\ref{appB}). Comparison of errors produced by such a state and a coherent state is presented in Fig.~\ref{fig2}b.

The numerical approach for  solving the eigenstates of $\hat{F}_{i}$ has the disadvantage of truncating the infinite-dimensional state vector to a finite vector of length $N_{\m{cut}}$, which might distort or exclude some of the possible solutions. However, the obtained Gaussian-like solutions are not affected by changes in the cut-off for $N_{\m{cut}}-\n\gg \sqrt{\n}$. Raising the cut-off reveals more energetic solutions, but these correspond to pulses that implement the chosen gate after an integer number of unnecessary $2\pi$ rotations.

Generally for $R'_{\theta}$ gates, we find solutions with errors that vanish as $1/\n$ in the limit $\n\rightarrow\infty$, as shown in Appendix~\ref{appC}. The lower bounds together with errors induced by non-squeezed coherent states are shown in Table~\ref{tab_1}. Other gates, such as the Pauli-Z gate and the Hadamard gate, can be constructed as sequences of $R'_{\theta}$ gates. Recently, it was shown that squeezing also improves the fidelity of the phase gate in the dispersive regime~\cite{Puri}.

\section{Drive-refreshing protocol \label{sec:protocol}}

\begin{figure}[tb!] \center	
	\includegraphics[width=0.9\linewidth]{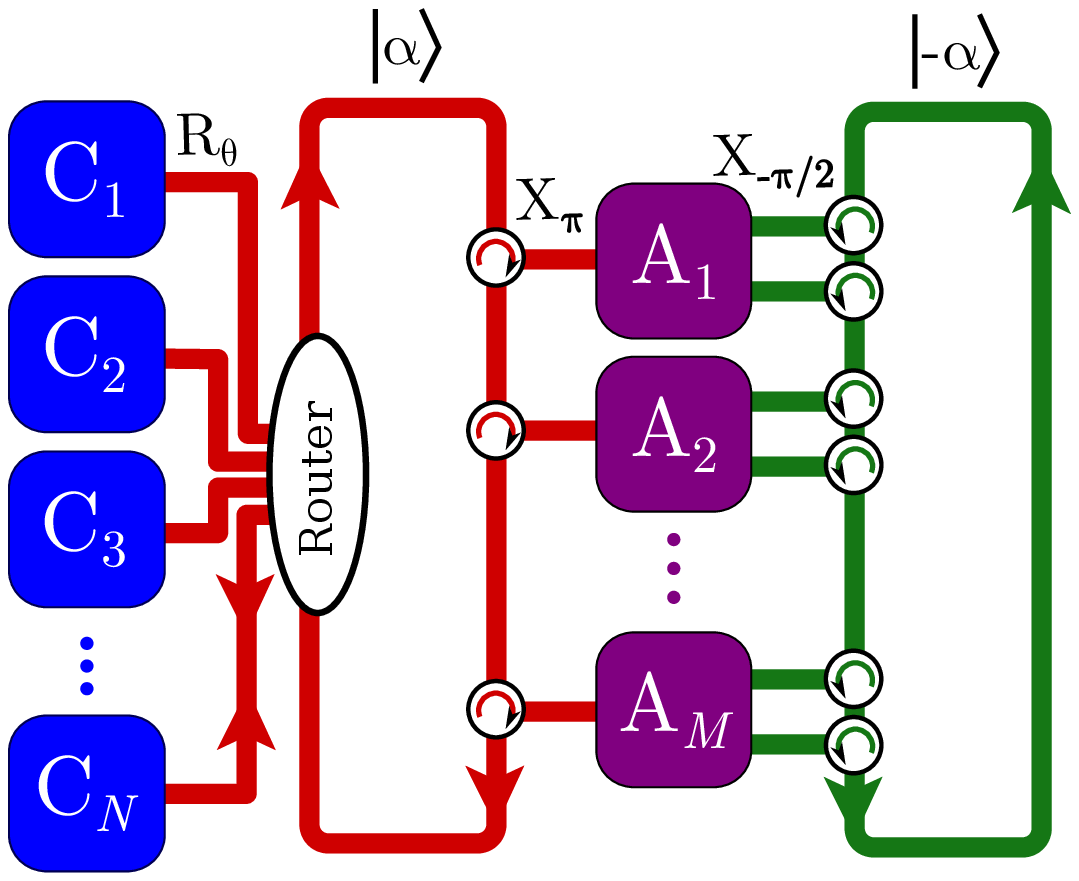}
	\caption{ \label{fig3} Schematic diagram of the drive-refreshing protocol. During one cycle, the circulating drive pulse (red) induces a chosen rotation $R_\theta$ on one of the qubits $C_{i}$ in the register and is then refreshed by sequential $X_{\pi}$ interactions with each ancillary qubit \{$A_{j}$\}. In an ideal setting, each ancilla is prepared precisely into the state $(\Ket{\m g_{\m A}}+i\Ket{\m e_{\m A}})/\sqrt{2}$ and reset after each cycle. In practice, the ancilla qubits are initially in their ground states and their preparation and reset is implemented by a circulating corrector pulse (green).
	 }% end caption
\end{figure}

All of the fundamental lower bounds derived above are inversely proportional to the average photon number. Intuitively, a drive with a large photon number should be capable of inducing multiple gates without changing substantially, thus decreasing the required amount of energy per gate for nearly equal error level. We show below that reusing a drive effectively decreases the energy consumption well below the lower bound of average gate error for disposable pulses. Furthermore, the drive can be corrected between successive gates such that the consumption drops without essential decrease of the average gate fidelity.

In our protocol, an itinerant control drive cyclically interacts with a register of resonant qubits and ancillary qubits, see Fig.~\ref{fig3}. A cycle begins with the drive, initially in a suitable squeezed coherent state, applying a chosen gate operation with minimal error on a register qubit. Consequently, the drive state changes due to the quantum back-action. To undo this, the drive is set to sequentially interact with corrective ancilla qubits, initialized in a superposition of ground and excited states, for a time corresponding to a $\pi$-rotation. As a result, the purity, energy, and phase of the drive are restored in successive interactions (see Appendix~\ref{appD}). At the end of the cycle, the ancilla qubits are reset and the refreshed drive is usable for another high-fidelity gate.

With increasing number of ancilla qubits, the execution time of a full cycle increases and thus   one itinerant pulse applies a gate on the register less frequently. To compensate for this, one could add another drive pulse for each ancilla in the array, and synchronize their travel times such that each qubit would interact with one of the pulses at a given time.  Such a system would apply as many gates on the register per cycle as there are itinerant pulses in circulation. However, we restrict our analysis to a single pulse.

\begin{figure}[tb!] \center	
	\includegraphics[width=0.7\linewidth]{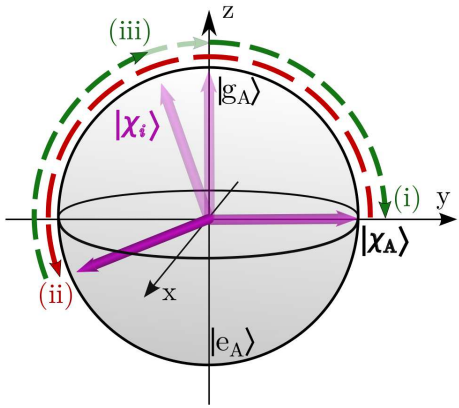}
	\caption{ \label{fig4} Evolution of an ancilla state during a refreshing cycle: (i) preparation from the ground state into the state $\Ket{\chi_{\m A}}=(\Ket{\m g_{\m A}}+i\Ket{\m e_{\m A}})/\sqrt{2}$, (ii) drive refresh as a result the primary rotation (red), and (iii) ancilla reset. We either assume that the preparative steps (i) and (iii) are ideal or induced by a corrector pulse as shown in Fig.~\ref{fig3}.
	 }% end caption
\end{figure}

The refreshing by the ancillary interactions is understood by considering the path traversed by the Bloch vector of the ancillary qubit, as illustrated in Fig.~\ref{fig4}. A drive lacking energy rotates the vector with smaller angular frequency, leaving the ancilla slightly biased towards the ground state and gaining energy in the process. Similarly, excessive energy in the drive is transferred to the ancilla due to rotating it closer to the excited state.

The Hilbert space of this system is formally a composite space of the Fock space $D$ of the drive and the two-level spaces $\{Q_{k}\}_k=\{C_{i},\, A_{j}\}_{i,j}$ of the register and ancilla qubits,
\begin{equation}
S=\left(\bigotimes_{k=1}^{N+M}Q_{k}\right)\otimes D=\left(\bigotimes_{i=1}^{N}C_{i}\right)\otimes\left(\bigotimes_{j=1}^{M}A_{j}\right)\otimes D.
\end{equation}
The drive only interacts with one qubit at a time and therefore each interaction can be calculated in the subspace of the relevant qubit and the drive, assuming the qubits are not correlated. After the interaction, the drive state is extracted by tracing over the associated qubit space. Namely, the $i$th iteration of the drive state is given by
\begin{equation}
\hat{\sigma}_{i+1}=\m{Tr}_{Q_{i}}\left[\hat{U}_i\left(\Ket{\chi_{i}} \negthickspace\Bra{\chi_{i}}\otimes\hat{\sigma}_{i}\right)\hat{U}^{\dagger}_i\right], 
\label{eq:NextState}
\end{equation} where $\hat{U}_i$ acts in the subspace of the drive and the $i$th qubit in the protocol sequences described in the following sections.

\subsection{Implementation with ideally prepared ancilla\label{idealprotocol}}

Consider first the case where the ancilla qubits are perfectly reset during each cycle, and the gate we wish to apply on each register qubit is $X_\pi$. The protocol is executed with the following steps:
\begin{enumerate}[label=(\roman*)]
\item The drive state is initialized to the $\overline\err$-minimizing state $\Ket{\sigma_{0}}=|\sqrt{\n},\sq\rangle$.
\item A new register qubit is initialized in a random pure state, chosen uniformly from the Bloch sphere.
\item The drive interacts with the register qubit for interaction time $T=\pi/(2g\n)$ {[}Eq.~\eqref{eq:NextState}{]}.
\item The $M$ ancilla qubits are initialized to $\Ket{\chi_{\m A}}=(\Ket{\m g_{\m A}}+i\Ket{\m e_{\m A}})/\sqrt{2}$.
\item The drive interacts with an ancilla qubit for interaction time $T=\pi/(2g\n)$. Repeat for all ancillas.
\item Evaluate the average error $\overline{\err}$ of a hypothetical $X_{\pi}$ gate with Eqs.~\eqref{eq:AverageError} and~\eqref{eq:errorTraces} using the current drive state. Continue from step~(ii).
\end{enumerate}
For gates other than $X_\pi$, the phases of the drive and ancillas, as well as the interaction time in step~(iii), but not step~(v), would be shifted accordingly.

\begin{figure}[tb] \center	
	\includegraphics[width=1\linewidth]{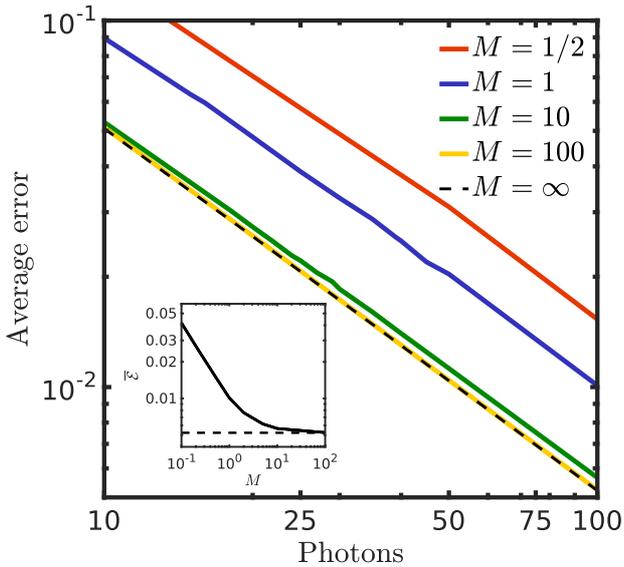}
	\caption{ \label{fig5} Average error $\overline{\err}$ of $X_{\pi}$ gates generated by an itinerant drive pulse which initially had an average photon number $\n$ and has reached the steady state due to ancilla refreshing. The drive is set to interact with $M$ ideal ancillas ($\Ket{\chi_{\m A}}=(\Ket{\m g_{\m A}}+i\Ket{\m e_{\m A}})/\sqrt{2}$) per cycle as indicated, leading to effective refreshing of the drive state. The dashed line indicates the lower bound of error which is achieved either with a disposable optimal pulse or with a pulse refreshed by infinitely many ideal ancillas. The inset shows the average gate error as a function of $M$ for $\n = 100$.
	 }
\end{figure}

We numerically simulate the evolution of the drive and evaluate the average error of the gate $X_{\pi}$ for a register qubit after each cycle. During the protocol, the average error $\overline{\err}$ will increase from its initial lower-bound value at varying rates depending on the randomized states of the register qubits. We find that after many cycles, the drive reaches a steady state that generates the desired gates with a predictable average error. With 1--3 ancillas per cycle, the average error saturates after a hundred cycles; with ten or more ancillas, the saturation takes less than ten cycles. If no corrective ancillas are used, the average error eventually reaches $0.5$.

Figure~\ref{fig5} shows how the eventual error level depends on the number of photons and ancillas. The average gate error approaches its theoretical lower bound, in the limit of many drive-refreshing ancilla qubits. For smaller rotation angles, qualitatively similar results are obtained with more slowly accumulating error. Thus a single itinerant drive pulse supplied with ideal ancilla states can generate an infinite number of high-fidelity gates.

\subsection{Register in an entangled state}

In the previous section, the qubits in the register were assumed to be essentially uncorrelated to justify the partial tracing over each qubit after the respective interaction. Here we demonstrate the beneficial performance of our method in the case where the register qubits are maximally entangled. 
We initialize the register of $N$ qubits in the Greenberger--Horne--Zeilinger (GHZ) state $\k{\psi_{\mathrm{GHZ}}}=(\k{\mathrm{g}}^{\otimes N}+\k{\mathrm{e}}^{\otimes N})/\sqrt{2}$. The control protocol is physically the same as in the previous section: the drive interacts with only one qubit at a time to implement a single-qubit gate $\hat{K}$ and is refreshed by $M$ ideally prepared ancillas between each such gate. The target operation on the register is thus $\hat{K}^{\otimes N}$. 
Due to the entangled register, the temporal evolution operators must be calculated in the Hilbert space $\left(\bigotimes_{i=1}^{N}C_{i}\right)\otimes D$ or $\left(\bigotimes_{i=1}^{N}C_{i}\right)\otimes A_j \otimes D$  for interactions between the drive and a register qubit, or drive and the $j$th ancilla, respectively. No partial trace over any register qubit is taken. After the drive has interacted with every register qubit once, the state of the register has transformed into $\hat{\rho}'$ and  the total transformation error is computed as 
\begin{equation}
\err_{\mathrm{GHZ}}=1-\m{Tr}\left[ \hat{\rho}'\hat{K}^{\otimes N}\kb{\psi_{\mathrm{GHZ}}}{\psi_{\mathrm{GHZ}}}(\hat{K}^{\dagger})^{\otimes N}\right].
\end{equation} 
We divide this error by the number of qubits to obtain the effective error per gate, $\err_{\mathrm{eff}}=\err_{\mathrm{GHZ}}/N$.

\begin{figure}[tb] \center	
	\includegraphics[width=1\linewidth]{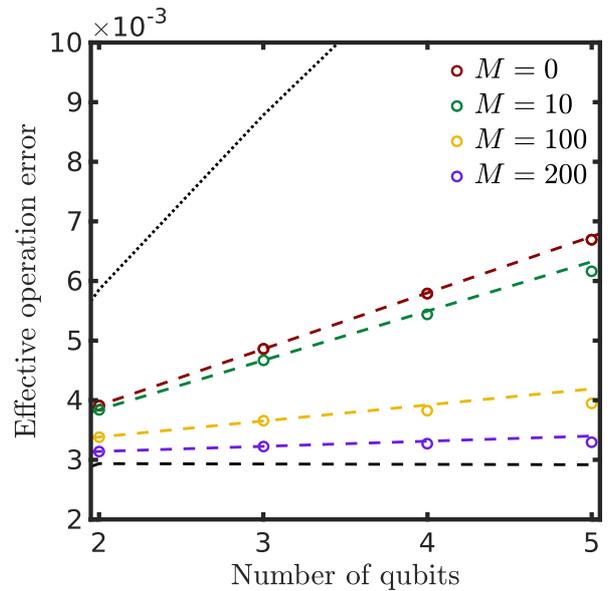}
	\caption{ \label{fig6} State preparation error per qubit $\err_{\mathrm{eff}}=\mathcal{E}_\mathrm{GHZ}/N$ for a register of $N$ qubits initially in a GHZ state. The target gate is an $X_{\pi/2}$ rotation for all qubits individually, implemented by a squeezed state of $\n=100$ photons ($r=\ln{\sqrt{\pi/2}}$) that is refreshed by $M$ ideal ancillas per cycle. The circles represent the data, whereas the coloured lines extend the line segments between the first two data points, to distinguish deviations from linear behaviour. The black dashed line represents the error obtained using $N$ disposable pulses of constant photon number $\n=100$. The dotted line is the error due to disposable pulses of constant total energy $\n=100/N$.
	 }
\end{figure}

Results of a simulation for an $X_{\pi/2}$ gate with the initial drive state $\Ket{\sqrt{100},\ln\sqrt{\pi/2}}$ are shown in Fig.~\ref{fig6}. A behaviour similar to Fig.~\ref{fig5} is observed: with enough ancillary corrections between the register gates, the error produced by an itinerant drive can be reduced to the level given by individual pulses. The figure also suggests that even without corrections, reusing a drive of certain energy is more beneficial in practice than dividing the same amount of  photons into individual, weaker disposable pulses.
Thus we conclude that regardless of the state of the register, refreshment of a drive pulse likely serves to improve the trade-off between gate error and required energy.

The above case of entangled qubits also provides a way to compare our results to the previous work by Gea-Banacloche and Ozawa~\cite{Ozawa2}, where they studied a register in a GHZ state that was operated by a drive of $\n$ photons on average. They showed that the maximum error of the $X_{\pi/2}$ gate in this system scales as $N/\n$ per qubit. This scaling was used to argue that a pulse of average photon number $\n_\mathrm{total}$ cannot outperform $N$ individual pulses of $\n_\mathrm{total}/N$ average photons, although their performance was not compared explicitly. The key differences here are that Ref.~\cite{Ozawa2} does not consider the possibility of using ancillary qubits, and that it employs a definition of error which also accounts for the infidelity of the drive state. Our results suggest that even though the errors due to both reused and disposable pulses of equal total energy increase almost linearly with $N$, the  prefactor of the former is much smaller and can be greatly improved by the refreshing protocol.
%negatiivinen etutekijä?

\subsection{Full protocol\label{fullprotocol}}

\begin{figure}[tb] \center	
	\includegraphics[width=1\linewidth]{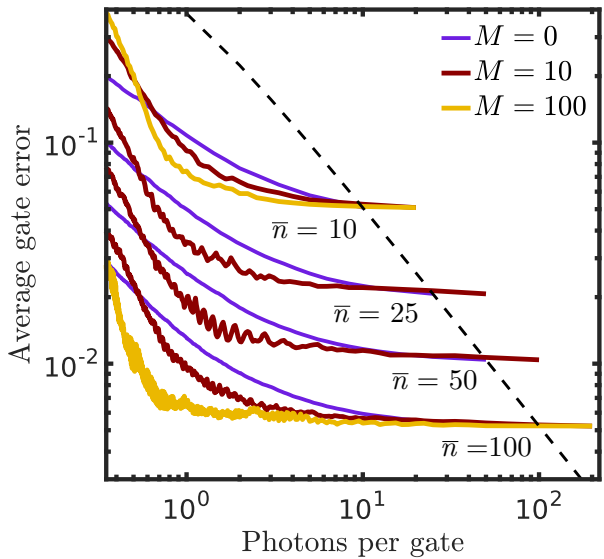}
	\caption{ \label{fig7} Average gate error as a function of the total mean number of initial photons, $2\n$ for $M>0$ and $\n$ for $M=0$, divided by the number of $X_{\pi}$ register gates generated. The ancilla states are non-ideally prepared by a corrector pulse initially in state $\Ket{-\sqrt{\n},\sq}$. During the protocol, the curve advances from right to left and the results are averaged over multiple simulations. The dashed line indicates the lower bound of error which is achieved either with a disposable optimal pulse or with a pulse refreshed by infinitely many ideal ancillas.
	 }
\end{figure}

The total energy consumption of the protocol can be meaningfully estimated only if the method and energy cost of the ancilla preparation is specified. To this end, we propose to prepare the ancillas by a circulating corrector pulse shown in Fig.~\ref{fig3}. In the full protocol, the ancilla qubits are first prepared in their ground state and then controlled by the corrector pulse from cycle to cycle. With opposite phase and half the interaction time compared with the drive, the corrector pulse applies an $X_{-\pi/2}$~gate on the ancilla before and after a $X_{\pi}$~gate introduced by the drive pulse. For simplicity, we assume that the state of the register is separable. The full protocol is given by the following steps: 
\begin{enumerate}[label=(\roman*)]
\item The drive state is initialized to $|\sqrt{\n},\sq\rangle$, the corrector pulse to $|-\sqrt{\n},\sq\rangle$ and all $M$ ancillas to the ground state.
\item A new register qubit is initialized in a random pure state. 
\item The drive interacts with the register qubit with interaction time $T=\pi/(2g\n)$ {[}Eq.~\eqref{eq:NextState}{]}. 
\item An ancilla qubit interacts sequentially with the corrector, the drive, and the corrector again, with interaction times $T/2$, $T$, and $T/2$, respectively.  Repeat for all other ancillas.
\item Evaluate the average error $\overline{\err}$ of a hypothetical $X_{\pi}$ gate with Eqs.~\eqref{eq:AverageError} and \eqref{eq:errorTraces} using the current drive state. Continue from step~(ii).

\end{enumerate}

In addition to computing the drive state after each interaction, the state of the interacting qubit is also extracted for subsequent use by a partial trace over the drive degrees of freedom. This is justified if the ancilla qubits do not become strongly correlated during the evolution. This approximation is more accurate the closer the control pulses are to classical pulses which do not induce entanglement.

Since all ancilla qubits are prepared to the ground state, the energy consumption fully arises from the drive and corrector pulses, both of which have the initial average energy $\n\hbar\omega$. Thus, the average energy consumption per register gate is $E=2\hbar\omega\n/N$, where $N$ is the number of elapsed cycles, or equally gates generated. In the case where the drive-refreshing protocol is not used, $M=0$, we have $E=\n\hbar\omega/N$.

Results from multiple simulations are averaged and shown in Fig.~\ref{fig7}. In contrast to the ideal case, the system accumulates error over repeated cycles and the average gate error does not saturate. Nevertheless, we find that with a sufficient number of ancillary qubit interactions between the register gates, the average error remains nearly constant for a large number of successive gates. The protocol can be stopped before the error reaches a desired threshold. This shows that the total energy cost per register gate is effectively reduced to orders of magnitude below the lower bound for disposable pulses. In fact, Fig.~\ref{fig7} suggests that the gate error may be, in theory, reduced indefinitely without increasing the power consumption by using more energetic pulses.

\section{Discussion \label{sec:discussion}}

In this work, we derived the greatest lower bound for the error of a single-qubit gate implemented with a single resonant control mode of certain mean energy. In contrast to previous work, our method for obtaining the bound is not restricted to any particular gate or state of the qubit--drive system. The method can also be used to find the quantum state  of the drive mode that minimizes the average gate error, or alternatively the transformation error for a chosen initial qubit state. Specifically, we found that the lower bounds for rotations about axes in the $xy$-plane are achieved by squeezing the quantum state of a coherent drive pulse by an amount that depends on the target gate. Together with the recent result that squeezing also significantly improves the phase gate in the dispersive regime~\cite{Puri}, our results suggest that squeezing may generally yield useful improvements in different control schemes. This calls for experimental studies on outperforming the widely-used coherent state.

Importantly, our results also impose a lower bound on the energy consumption of individually driven qubits. Delivering the required power to the qubit level, possibly through a series of attenuators, implies heat management challenges that must be addressed in future large-scale quantum computers. As a solution, we introduced a concrete protocol where an itinerant control pulse is used to generate multiple gates and is refreshed between them to avoid loss of gate fidelity. The refreshing process may also prove useful in correcting the phase and amplitude errors of a noisy control pulse.

Our protocol can possibly be realized in some form with future low-loss microwave components such as photon routers~\cite{Pechal,Hoi}, circulators, and nanoelectromechanical systems~\cite{Zhou}. Technical limitations in the quality of these devices will set in practice the trade-off between the achievable gate fidelity and the dissipated power. In the future, our work can be extended to error bounds for 2-qubit gates, state preservation, pulse amplification, and propagating control pulses composed of a continuum of bosonic modes.

\begin{acknowledgments}

We thank Paolo Solinas and Benjamin Huard for useful discussions. This work was supported by the European Research Council under Starting Independent Researcher Grant No. 278117 (SINGLEOUT) and under Consolidator Grant No. 681311 (QUESS). We also acknowledge funding from the Academy of Finland through its Centres of Excellence Program (grant nos\ 251748 and 284621) and grant (no.\ 286215) and from the Finnish Cultural Foundation.

\end{acknowledgments}

\renewcommand{\appendixname}{APPENDIX}
\appendix

\section{ESTIMATED ENERGY CONSUMPTION OF A SURFACE CODE \label{appA}}

We estimate the power required by a superconductor-based quantum computer solving a 2000-bit factorization problem, stabilized by a surface code. For this particular computation, the needed number of physical qubits has been estimated by Fowler~\cite{Fowler} to be $N_{\m q}\approx2\times 10^{8}$. We assume that the physical qubits are controlled with typical coherent microwave pulses and that $\pi/2$ gates are completed in equal time and with lower power than $\pi$ gates. The average power needed during one surface code cycle is calculated by counting the frequency of measurements, $R_\pi$, $R_{\pi/2}$, and CNOT operations, and by taking a duration-weighted average of the corresponding powers. The  operation times depend on implementation. Using operation times achieved in Ref.~\cite{Kelly}, $T_{\pi}=20$ ns, $T_{\m{CZ}}=45$ ns, and $T_{\m{M}}=140$ ns, for $\pi$-rotations, controlled phase gates, and measurements, respectively, and assuming that our code executes as many operations in parallel as possible, the average power per physical qubit is approximately $P_\m{q}=\frac{15}{184}P_\pi+ \frac{9}{46}P_{\m{CZ}}+ \frac{7}{92}P_{\m{M}}$ where the $P$'s denote the average drive powers for the the above-mentioned operations. For simplicity, we neglect the two-qubit gates and measurements and use $P_{\m{q}}=\frac{15}{184}P_\pi\approx0.1P_\pi$.

Typical powers at the chip are of the order of $P_\pi=10^{-11}$ W, after being generated in the room temperature and attenuated by tens of decibels on their way to roughly 10-mK base temperature. Using only $10$~dB of attenuation at the base temperature, the total power dissipation here becomes $P= N_{\m q}P_\m{q}\times 10 \approx 2$~mW. Such power level is much higher than the typical cooling power of $10~\mu$W in state-of-the-art dilution refrigerators at 10~mK.

Note that using an open transmission line is expected to consume more power than required in the single-mode case considered in Sec.~\ref{sec:optimization}. The average energy density in a transmission line is given by $E_l = \mathcal{C}V_{\m{rms}}^2$, where $\mathcal{C}$ is the capacitance per unit length and $V_{\m{rms}}$ is the root mean square of the voltage. In a time interval $T_\pi\gg\omega^{-1}$, a propagating drive pulse advances a distance $l=\frac{\lambda\omega}{2\pi}T_\pi$, effectively transporting a power of $P_\pi^\m{TL}=E_l l/T_\pi=\frac{1}{2\pi}\mathcal{C}\lambda V_{\m{rms}}^2 \omega$, where $\lambda$ is the photon wavelength. In comparison, consider a $\lambda/4$ resonator which is used to apply $V_{\m{rms}}$ to the qubit for an equal operation time. The resonator requires a power $P_\pi^\m{R}=\frac{1}{8}\mathcal{C}\lambda V_{\m{rms}}^2/T_\pi$, and with a typical qubit frequency of $\frac{\omega}{2\pi}=6$ GHz, the ratio between the powers is $P_\pi^\m{TL}/P_\pi^\m{R}=\frac{4}{\pi}\omega T_\pi\approx150$. 
Thus qubit control using propagating photons in a transmission line seems to lead to orders of magnitude higher power consumption than our single-mode case. However,
a more comprehensive study employing the quantization of the transmission line is required to reach accurate estimates. We leave such study for future research.

Finally, let us consider the lower bound for the power to drive the qubits using disposable pulses. The minimum amount of photons (see Sec.~\ref{minimizationmethod}) to produce the gate error $\overline{\err}=0.1\%$  used by Fowler in Ref.~\cite{Fowler} is $\n=\frac{4+\pi^{2}}{24\overline{\err}}\approx500$ photons at the qubit level. With $\frac{\omega}{2\pi}=6$ GHz, the corresponding powers are $P_{\pi,\min}=\hbar\omega\n/T_{\pi}\approx 10^{-13}$ W and $P_{\min} \approx 20~\mu$W. This suggests that the lower bound for our example problem size is at the border where current refrigeration equipment fail to deliver the required cooling power, and hence significant increments in the problem size or non-ideal implementation of the suggested driving techniques call for inventive solutions to the emerging heat management problem.

A way to avoid the attenuation at the base temperature would be to generate the control pulses at the chip level. To our knowledge, however, no present chip-level photon source is capable of producing pulses that are accurate and intense enough to induce quantum gates of high fidelity. Furthermore, the operation efficiency of such devices needs to be sufficiently high to be a considerable alternative. Typically microwave sources internally dissipate much more power than their maximum output.

\section{OPTIMIZATION OF THE GATE ERROR \label{appB}}
\subsection{Gate error for a given initial qubit state}

Assuming the qubit--drive system is initially in a pure state, $\c_{0}\otimes\s_{0}\equiv\k{\chi_{0}}\negthickspace\b{\chi_{0}}\otimes\k{\sigma_{0}}\negthickspace\b{\sigma_{0}}\equiv\kb{\chi_{0},\sigma_{0}}{\chi_{0},\sigma_{0}}$, Eq.~\eqref{eq:errorTraces} reduces to
\begin{gather}
\err(\chi_{0},\sigma_{0})=1-\m{Tr}\left\{ \m{Tr}_{D}\left[\hat{U}(T)\left(\c_{0}\otimes\s_{0}\right)\hat{U}^{\dagger}(T)\right]\hat{K}\c_{0}\hat{K}^{\dagger}\right\} \nonumber \\
\quad\;=1-\sum_{k=0}^{\infty}\left|\b{\chi_{0},k}(\hat{K}^{\dagger}\otimes\hat{\mathbb{I}})\hat{U}(T)\k{\chi_{0},\sigma_{0}}\right|^{2}, \label{eq:err1}
\end{gather}
where $\hat{K}$ is the desired gate, $T$ is the interaction time, $\hat{U}(T)$ is the temporal evolution operator, and $\{\Ket{k}\}$ are the photon number states. We represent the basis of the qubit space using vectors $\k{\m g}\widehat{=}\left(\begin{array}{c}
1\\
0
\end{array}\right)$ and $\k{\m e}\widehat{=}\left(\begin{array}{c}
0\\
1
\end{array}\right)$, and explicitly write
\begin{gather}
\k{\sigma_{0}}=\sum_{j=0}^{\infty}c_{j}\k j\negmedspace,\label{eq:drive1}\\
\k{\chi_{0}}\widehat{=}\left(\begin{array}{c}
\cos\frac{\vartheta}{2}\\
\sin\frac{\vartheta}{2}\e{i\varphi}
\end{array}\right). \label{eq:qubit}
\end{gather}
In this basis, $\hat{K}\widehat{=}\left(\begin{array}{cc}
K_{11} & K_{12}\\
K_{21} & K_{22}
\end{array}\right)$, and $\hat{U}(T)$ is given by
\begin{equation}
\hat{U}(T)\,\widehat{=}\,\sum_{l=0}^{\infty}\left(\begin{array}{cc}
C_{l}(T)\kb ll & -iS_{l+1}(T)\kb{l+1}l\\
-iS_{l}(T)\kb{l-1}l & C_{l+1}(T)\kb ll
\end{array}\right),
\end{equation}
with the shorthand notations $C_{k}(T)=\cos\left(gT\sqrt{k}\right)$, $S_{k}(T)=\sin\left(gT\sqrt{k}\right)$, and $|-1\rangle=0$. Using the expressions above, the matrix element in Eq.~\eqref{eq:err1} can be structured as
\begin{equation}
\b{\chi_{0},k}(\hat{K}^{\dagger}\otimes\hat{\mathbb{I}})\hat{U}(T)\k{\chi_{0},\sigma_{0}} = \sum_{mn=0}^{1}\Gamma_{nm}^{k}c_{k+n-m}, \label{eq:Gee}
\end{equation}
where
\begin{eqnarray}
\Gamma_{00}^{k}(\vartheta,\varphi) & = & C_{k}\left[K_{11}^{*}\cos^{2}\left(\frac{\vartheta}{2}\right)+K_{12}^{*}\frac{1}{2}\sin\left(\vartheta\right)\e{-i\varphi}\right],\nonumber \\
\Gamma_{01}^{k}(\vartheta,\varphi) & = & -iS_{k}\left[K_{12}^{*}\sin^{2}\left(\frac{\vartheta}{2}\right)+K_{11}^{*}\frac{1}{2}\sin\left(\vartheta\right)\e{i\varphi}\right],\nonumber \\
\Gamma_{10}^{k}(\vartheta,\varphi) & = & -iS_{k+1}\left[K_{21}^{*}\cos^{2}\left(\frac{\vartheta}{2}\right)+K_{22}^{*}\frac{1}{2}\sin\left(\vartheta\right)\e{-i\varphi}\right],\nonumber \\
\Gamma_{11}^{k}(\vartheta,\varphi) & = & C_{k+1}\left[K_{22}^{*}\sin^{2}\left(\frac{\vartheta}{2}\right)+K_{21}^{*}\frac{1}{2}\sin\left(\vartheta\right)\e{i\varphi}\right].\nonumber
\end{eqnarray}
The error is thus given by
\begin{equation}
\err(\vartheta,\varphi)=1-\sum_{k=0}^{\infty}\left|\sum_{m,n=0}^{1}\Gamma_{nm}^{k}(\vartheta,\varphi)c_{k+n-m}\right|^{2}.\label{eq:err2}
\end{equation}

We can define the transformation operator through its matrix elements in the photon number basis as $\hat{F}(\vartheta,\varphi)=\sum_{i,j=0}^{\infty}F_{ij}(\vartheta,\varphi)\kb ij$, where
\begin{equation}
F_{ij}(\vartheta,\varphi) = \sum_{k=0}^{\infty}\sum_{\substack{ m',n', \\ m,n=0}}^{1}\Gamma_{nm}^{k}\left(\Gamma_{n'm'}^{k}\right)^{*}\delta_{i+m'-n',j+m-n\geq0}, \label{eq:Amax}
\end{equation}
and $\delta_{ij\geq0}$ is a Kronecker delta that is zero for any negative index. Equation~\eqref{eq:err2} is thus reduced to the form given in Eq.~\eqref{eq:generalCompact}, that is,
\begin{eqnarray}
\err(\vartheta,\varphi) & = & 1-\sum_{j,i=0}^{\infty}F_{ij}(\vartheta,\varphi)c_{i}^{*}c_{j} \nonumber \\
& = & 1-\b{\sigma_{0}}\hat{F}(\vartheta,\varphi)\k{\sigma_{0}}.
\end{eqnarray}

\subsection{Average gate error}

Using Eq.~\eqref{eq:AverageError}, the average error and its corresponding operator can be structured in a similar manner. Defining the matrix elements of the operator $\hat{F}_{\m{avg}}$ as
\begin{eqnarray}
F_{\m{avg},ij} & = & \sum_{k=0}^{\infty}\sum_{\substack{ m',n', \\ m,n=0}}^{1}\frac{1}{4\pi}\int_{0}^{\pi}\int_{0}^{2\pi}\Gamma_{nm}^{k}\left(\Gamma_{n'm'}^{k}\right)^{*} \nonumber \\
& &  \times\sin\vartheta\,\m d\vartheta\,\m d\varphi\,\delta_{i+m'-n',j+m-n\geq0}, \label{eq:Amatrix}
\end{eqnarray}
the average error also assumes the form of Eq.~\eqref{eq:generalCompact}.

As shown in Ref.~\cite{Bowdrey}, the average error integrated over the Bloch sphere is equal to the arithmetic mean of the error of six so-called axial states. This provides an alternative expression for the operator $\hat{F}_{\m{avg}}$, namely,
\begin{eqnarray}
\hat{F}_{\m{avg}} & = & \frac{1}{6}\left[\hat{F}\left(0,0\right)+\hat{F}\left(\pi,0\right) + \hat{F}\left(\frac{\pi}{2},0\right)\right. \nonumber \\
& & \left.+\hat{F}\left(\frac{\pi}{2},\pi\right) + \hat{F}\left(\frac{\pi}{2},\frac{\pi}{2}\right) + \hat{F}\left(\frac{\pi}{2},-\frac{\pi}{2}\right)\right].
\end{eqnarray}

\subsection{Maximum gate error}

We can optimize the maximum error if there exists an initial qubit state which produces the highest error regardless of the drive state, i.e., $\err_{\max}=\err(\vartheta_{s},\varphi_{s}) = \max_{\vartheta,\varphi}\err(\vartheta,\varphi)$. Specifically for gates $R'_{\pi}$, computing the gradients of $\err$ with respect to $\vartheta$ and $\varphi$ shows that the maximum point is virtually independent of the drive state, and that the maximum error is obtained with $\vartheta_{s}=\frac{\pi}{2}$ and $\varphi_{s}=\phi\pm\frac{\pi}{2}$, or equivalently $\k{\chi_{s}}=\frac{1}{\sqrt{2}}\left(\k{\m g}\pm i\e{i\phi}\k{\m e}\right)$, where $\phi$ is the angle between the horizontal rotation axis $R'$ and the $x$-axis. Due to symmetry, the initial drive state that optimizes $\err_{\max}$ is an eigenvector of
\begin{equation}
\hat{F}_{\min}=\frac{1}{2}\hat{F}\left(\frac{\pi}{2},\phi+\frac{\pi}{2}\right)+\frac{1}{2}\hat{F}\left(\frac{\pi}{2},\phi-\frac{\pi}{2}\right), \label{eq:fmin}
\end{equation}
which corresponds to the mean error of these two states.

The elements of the commutator $[\hat{F}_{\min},\hat{F}_{\m{avg}}]$ turn out to decrease as $\n\,^{-3}$. Thus it follows that the eigenvectors of $\hat{F}_{\m{avg}}$, i.e., the squeezed cat states given by Eq.~\eqref{eq:Optimal state}, simultaneously minimize both $\err_{\max}$ and $\overline{\err}$ in the limit $\n\rightarrow\infty$.

\section{APPROXIMATE GATE ERROR \label{appC}}

This section shows how the gate error can be analytically approximated for a specific gate. As an example, we choose $\overline{\err}$ for the gate $X_{\pi}$. Using $\hat{K}=\hat{\sigma}_{x}$, it is straightforward to evaluate the integrals in Eq.~\eqref{eq:Amatrix}, and hence the average error {[}Eq.~\eqref{eq:generalCompact}{]} becomes
\begin{eqnarray}
\overline{\err}(\sigma_{0},T) & = & \frac{2}{3}-\frac{1}{6}\sum_{n=0}^{\infty}\left\{ \abs{c_{n}}^{2}\left[S_{n}^{2}(T)+S_{n+1}^{2}(T)\right]\right. \nonumber \\
 & & \left. +2S_{n}(T)S_{n+1}(T)\m{Re}\left(c_{n+1}c_{n-1}^{*}\right)\right\}. \label{eq:ExactAverageError}
\end{eqnarray}
The error is then obtained by inserting the coefficients $\{c_{k}\}$ of the desired drive state: coherent, squeezed cat, or some other state. We choose a squeezed cat state
\begin{equation}
\k{\Sigma(\alpha,r)}=\frac{1}{\mathcal{N}}\left(\k{\alpha,r}\pm\k{-\alpha,r}\right),
\end{equation}
where $r$ is an unknown squeezing parameter and the amplitude satisfying $\alpha=\pi/(2gT)$ is real. In the high energy limit, the occupation numbers in a squeezed state $\k{\alpha,r}$ obey the normal distribution
\begin{equation}
\abs{c_{n}^{\m{squeezed}}(\alpha,r)}^{2}\approx\frac{1}{\sqrt{2\pi}\nu}\exp\left[-\frac{(n-\n)^{2}}{2\nu^{2}}\right],
\end{equation}
with mean $\n\approx\alpha^{2}$ and standard deviation $\nu\approx\alpha\e{-r}$. The amplitudes of a squeezed cat state $\k{\Sigma(\alpha,r)}=\sum_{n}\tilde{c}_{n}(\alpha,r)\k n$, are obtained by allowing either even (or odd) states to be occupied, such that
\begin{equation}
\tilde{c}_{n}(\alpha,r)\approx\sqrt{\frac{2}{\sqrt{2\pi}\e{-r}\alpha}} \exp\left[-\frac{\e{2r}}{4}\left(\frac{n}{\alpha}-\alpha\right){}^{2}\right],
\end{equation}
if $n$ is even (odd) and $\tilde{c}_{n}(\alpha,r)=0$ otherwise.

Insertion of these coefficients into Eq.~\eqref{eq:ExactAverageError} yields
\begin{equation}
\overline{\err}(\k{\Sigma(\alpha,r)})\approx\frac{2}{3}-\frac{S_{1}+S_{2}}{3\left(2\pi\e{-2r}\right)^{1/2}},
\end{equation}
where
\begin{widetext}
\begin{equation}
S_{1} =  \sum_{m=0}^{\infty}\alpha^{-1}\exp\left[-\frac{\e{2r}}{2}\left(\frac{2m}{\alpha}-\alpha\right)^{2}\right] \left[\sin^{2}\left(\frac{\pi}{2}\sqrt{2m\alpha^{-2}}\right)+\sin^{2}\left(\frac{\pi}{2}\sqrt{\left(2m+1\right)\alpha^{-2}}\right)\right],
\end{equation}
and
\begin{equation}
S_{2}  = \sum_{m=0}^{\infty}2\alpha^{-1}\exp\left(-\frac{\e{2r}}{2}\left\{ \left[\left(2m+2\right)\alpha^{-1}-\alpha\right]^{2}+\left(2m\alpha^{-1}-\alpha\right)^{2}\right\} \right) \sin\left[\frac{\pi}{2}\sqrt{\left(2m+1\right)\alpha^{-2}}\right]\sin\left[\frac{\pi}{2}\sqrt{\left(2m+2\right)\alpha^{-2}}\right].
\end{equation}
\end{widetext}
The sums can be computed by a change of variables $s\equiv\frac{2m}{\alpha}-\alpha$ and treating the infinite sum $\sum_{m=0}^{\infty}$ as an integral $\int_{-\infty}^{\infty}\frac{\alpha\m ds}{2}$, which justified in the limit $\alpha\rightarrow\infty$, where the functions are rather smooth and have support on a region much wider than unity. Approximating the result to the lowest order in $\n\,^{-1}$ eventually yields
\begin{equation}
\overline{\err}(\k{\Sigma(\alpha,r)})=\frac{4\e{2r}+\pi^{2}\e{-2r}}{24\alpha^{2}}+\mathcal{O}(\alpha^{-4}).
\end{equation}
This expression is minimized with the choice of the squeezing parameter $r=\ln\sqrt{\pi/2}$, independently of $\alpha$. Thus we have $\min_{\{c_{k}\}}\overline{\err}(\n)\approx\pi/(6\n)$.

Approximate average errors for gates $X_{\pi/2}$, $Y_{\pi}$, and $Y_{\pi/2}$ implemented by different drive states, such as the coherent state, can be computed in a similar fashion. This approach also works with the maximum error for rotations of $\pi$. Expressions obtained this way are listed in Table~\ref{tab_1}.

\section{FINDING THE STABILIZING ANCILLA STATE \label{appD}}

To build our protocol, we first search for initial states of the ancillas and the drive, such that each ancilla--drive interaction would steer the drive towards a stable state. For the protocol to work, the following conditions must be satisfied:
\begin{enumerate}[label=(\roman*)]
\item In the vicinity of the stable state, the drive is able to induce high-fidelity gates on a register qubit.
\item An ancilla--drive interaction increases the purity of the drive state, defined here as~\cite{Nielsen} $P(\hat{\sigma})=2\m{Tr}\left(\hat{\sigma}^{2}\right)-1$. Interactions with register qubits in randomized states tend to decrease the purity of a drive, rendering it less useful for subsequent gates. Thus increasing the purity effectively transfers entropy from the drive to the ancilla qubits.
\item An ancilla--drive interaction steers the amplitude, or equally energy, of the drive towards its steady state value. This is needed since we want to generate gates with a fixed interaction time.
\item The interaction steers the relative phase of the drive towards its initial value.
\end{enumerate}

\begin{figure}[tb!] \center
	\includegraphics[width=0.9\linewidth]{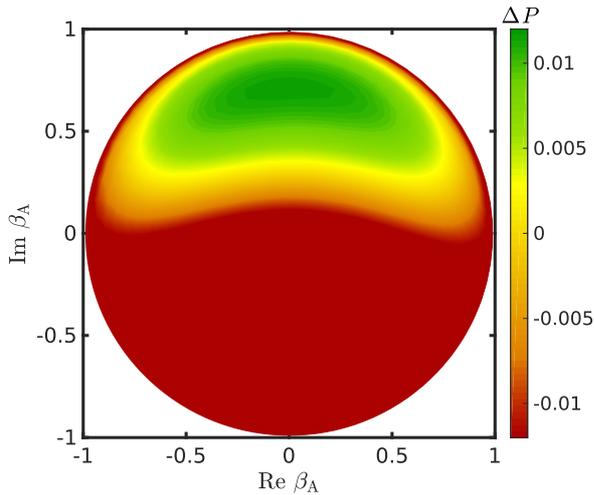}
	\caption{ \label{fig8} Purity change of a drive state due to the interaction with an ancilla qubit initialized in state $\protect\k{\chi_{\protect\m A}}=\sqrt{1-\protect\abs{\beta_{\protect\m A}}^{2}}\protect\k{\protect\m{g_{A}}}+\beta_{\protect\m A}\protect\k{\protect\m{e_{A}}}$. The shown change is $\Delta P_{10}=P(\hat{\sigma}_{11})-P(\hat{\sigma}_{10})$ averaged over multiple simulations, where $\hat{\sigma}_{11}$ is the result of the evolution of $\protect\kb{\chi_{\protect\m A}}{\chi_{\protect\m A}}\otimes\hat{\sigma}_{10}$. The impure initial drive state $\hat{\sigma}_{10}$ was obtained by letting a squeezed coherent state $\hat{\sigma}_{0}$ with $\protect\n=25$ interact with 10 qubits in randomly chosen initial states. The interaction time corresponds to a $\pi$ rotation.
}
\end{figure}

Condition~(i) suggests that the most promising candidates for the initial drive state are the coherent state $\Ket{\sqrt{\n}}$ and its squeezed variant $|\sqrt{\n},r\rangle$. We study condition~(ii) by computing the change in purity
\begin{equation}
\Delta P{}_{i}=P(\hat{\sigma}_{i+1})-P(\hat{\sigma}_{i})=2\m{Tr}_{D}\left(\hat{\sigma}_{i+1}^{2}-\hat{\sigma}_{i}^{2}\right),
\end{equation}
where $\hat{\sigma}_{i}$ is given by Eq.~\eqref{eq:NextState}. Figure~\ref{fig8} shows that the purity of the drive state increases if the ancilla is initialized close to the state $\Ket{\chi_{\m A}}=(\Ket{\m g_{\m A}}+i\Ket{\m e_{\m A}})/\sqrt{2}$. The figure also implies that the protocol works even if there is some error when preparing the ancilla in this state. 

\begin{figure}[tb!] \center
	\includegraphics[width=0.9\linewidth]{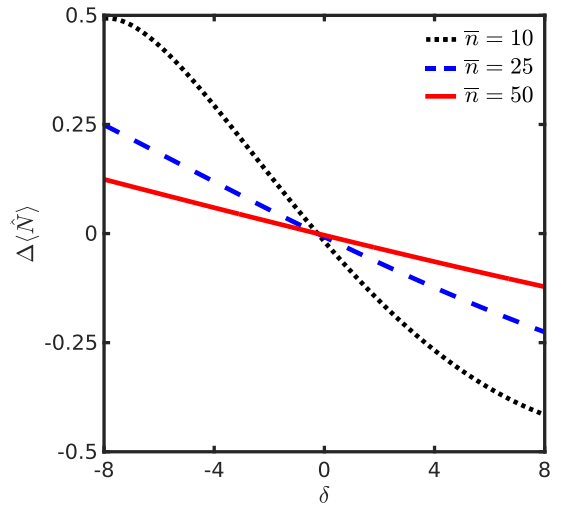}
	\caption{ \label{fig9} Change of the average photon number $\Delta\langle\hat{N}\rangle$ in an initial drive state $\protect\k{\sqrt{\protect\n+\delta},\ln\sqrt{\pi/2}}$ as a function of the deviation $\delta$. The change results from an interaction of a fixed time $T=\pi/\left(2g\sqrt{\protect\n}\right)$ with an ancilla qubit prepared in the state $\protect\k{\chi_{\protect\m A}}=\frac{1}{\sqrt{2}}\left(|\protect\m{g_{A}}\rangle+i|\protect\m{e_{A}}\rangle\right)$.
}
\end{figure}

Furthermore, changes in photon occupation,
\begin{equation}
\Delta\langle\hat{N}\rangle_{i}=\langle\hat{N}\rangle_{i+1}- \langle\hat{N}\rangle_{i}=\m{Tr}_{D}\left(\hat{a}^{\dagger} \hat{a}\hat{\sigma}_{i+1}-\hat{a}^{\dagger}\hat{a}\hat{\sigma}_{i}\right),
\end{equation}
with different initial energies of the drive shows in Fig.~\ref{fig9} that this ancilla state also enforces negative feedback on the average photon number, satisfying condition~(iii). Our study of the Wigner representation of the drive state after successive ancilla interactions shows that condition~(iv) is also satisfied.

For a protocol generating only $R_\pi$ gates, the squeezed cat state $|\Sigma(\sqrt{\n})\rangle$ was also tested as an initial drive state. Unfortunately, in this case conditions (ii) and (iii) do not hold for any choice of the ancilla state since the cat state does not rotate the Bloch vector of the ancilla in a specific direction, which is essential for the feedback mechanism depicted in Fig.~\ref{fig4}.

\FloatBarrier

% x words in appendices

%
%\bibliography{Refs}
%merlin.mbs apsrev4-1.bst 2010-07-25 4.21a (PWD, AO, DPC) hacked
%Control: key (0)
%Control: author (0) dotless jnrlst
%Control: editor formatted (1) identically to author
%Control: production of article title (0) allowed
%Control: page (1) range
%Control: year (0) verbatim
%Control: production of eprint (0) enabled
%

%x words in captions

\end{document}